\documentclass[10pt,twocolumn]{article}
\usepackage[utf8]{inputenc}
\usepackage[english]{babel}
\usepackage[letterpaper,top=2cm,bottom=2cm,left=2cm,right=2cm,columnsep=6mm]{geometry}
\usepackage{amsmath}
\usepackage{graphicx}
\usepackage{hyperref}
\usepackage{float}
\usepackage{times} 
\usepackage{multicol}
\usepackage{xcolor} 
\usepackage{caption}   
\usepackage{cuted}    
\usepackage{placeins}  
\hypersetup{
    colorlinks=true,
    linkcolor=blue,
    citecolor=blue,
    urlcolor=blue
}
\title{GeoPep: A geometry-aware masked language model for protein-peptide binding site prediction}
\author{
Dian Chen\textsuperscript{1}, 
Yunkai Chen\textsuperscript{2}, 
Tong Lin\textsuperscript{3,*}, 
Sijie Chen\textsuperscript{2}, 
Xiaolin Cheng\textsuperscript{2}
}

\begin{document}

\onecolumn  

\maketitle

\noindent
\textsuperscript{1}Department of Biomedical Engineering, Johns Hopkins University, Baltimore, Maryland, USA\\
\textsuperscript{2}College of Pharmacy, The Ohio State University, 281 W Lane Ave, Columbus, OH, USA\\
\textsuperscript{3}Amazon; \textsuperscript{*}This work is unrelated to work at Amazon.


\twocolumn[
\begin{@twocolumnfalse}
\maketitle
\begin{abstract}
\section{Abstract}

Multimodal approaches that integrate protein structure and sequence have achieved remarkable success in protein-protein interface prediction. However, extending these methods to protein-peptide interactions remains challenging due to the inherent conformational flexibility of peptides and the limited availability of structural data that hinder direct training of structure-aware models. To address these limitations, we introduce GeoPep, a novel framework for peptide binding site prediction that leverages transfer learning from ESM3, a multimodal protein foundation model. GeoPep fine-tunes ESM3's rich pre-learned representations from protein-protein binding to address the limited availability of protein-peptide binding data. The fine-tuned model is further integrated with a parameter-efficient neural network architecture capable of learning complex patterns from sparse data. Furthermore, the model is trained using distance-based loss functions that exploit 3D structural information to enhance binding site prediction. Comprehensive evaluations demonstrate that GeoPep significantly outperforms existing methods in protein-peptide binding site prediction by effectively capturing sparse and heterogeneous binding patterns. 

\end{abstract}
\vspace{1cm}
\end{@twocolumnfalse}
]

\clearpage
\section{Introduction}
Protein–peptide interactions underpin diverse cellular processes, including signal transduction, immune recognition, and transcriptional regulation, and constitute a major class of therapeutic targets. Peptides offer a unique pharmacological profile, combining the structural adaptability of small molecules with the specificity of biologics. Unlike small-molecule inhibitors, which require well-defined binding pockets and often exhibit off-target promiscuity, peptides can engage extended protein surfaces through multivalent contacts. At the same time, they circumvent key limitations of biologics, such as poor tissue penetration, immunogenicity, and manufacturing complexity. These properties render peptides promising candidates for modulating the estimated 80\% of disease-associated proteins that remain inaccessible to conventional drug modalities.

Despite their therapeutic potential, structural characterization of protein–peptide complexes is challenging~\cite{tsaban2022harnessing,johansson2019predicting}. These interactions are frequently transient and involve substantial conformational rearrangements, resulting in scarce high-resolution structural data for protein-peptide complexes~\cite{yin2024leveraging,lei2021deep}. The limited availability of experimental data necessitates computational methods to predict peptide binding sites and poses and quantify interaction energetics~\cite{wang2022predicting,yuan2024genome}.

Traditional physics-based strategies—such as molecular docking and molecular dynamics (MD) simulation—have been widely applied~\cite{chen2024design, weng2020comprehensive} but face fundamental limitations: (1) the inherent conformational flexibility of peptides necessitates exhaustive sampling beyond feasible timescales~\cite{wang2019improved}; (2) the absence of experimentally determined peptide structures complicates initial pose generation~\cite{chen2024design}; (3) inadequate representation of solvation effects introduces systematic errors in binding affinity estimation~\cite{chen2024design}; and (4) scoring functions lack discriminative power for native-like conformations versus decoys~\cite{weng2020comprehensive}. These limitations have motivated the development of machine learning-based approaches.


Deep learning models such as ScanNet and PeSTo have demonstrated strong performance in protein-protein interface prediction\cite{tubiana2022scannet, krapp2023pesto}. However, they fail to generalize to peptide-protein interactions, which exhibit sparse, localized contact patterns rather than extensive interfacial networks. Large-scale protein structure prediction models, including AlphaFold, show variable performance on peptide-protein complexes, with accuracy strongly dependent  on the presence of characterized binding motifs. AlphaFold2 achieves 75\% accuracy for motif-containing sequences versus 36\% accuracy for non-motif interactions \cite{tsaban2022harnessing}. The performance of specialized models such as PepNN remains limited due to the paucity of peptide-protein complex structures for training\cite{abdin2022pepnn}.


To address these limitations, we introduce GeoPep, a geometry-aware peptide binding site prediction model that combines transfer learning from protein foundation models with parameter-efficient complexity modeling. GeoPep fine-tunes ESM3, a multimodal protein foundation model trained on large-scale structural and sequence data, for peptide binding prediction. This strategy transfers ESM3's pre-learned protein representations to the peptide domain to overcome the data scarcity limitation in modeling peptide-protein interaction. To capture the sparse and heterogeneous interaction patterns characteristic of peptide binding sites, we employ Kolmogorov-Arnold Networks (KANs) \cite{liu2024kan} as parameter-efficient downstream predictors, which can capture intricate complex relationships from limited data.  Furthermore, our framework incorporates distance-based loss functions operating at the structural level to ensure geometric consistency in binding site prediction. Benchmarking across diverse datasets demonstrates that GeoPep achieves state-of-the-art performance, underscoring the utility of foundation model adaptation for peptide-centric prediction tasks.


\section{Results}

\subsection{GeoPep architecture for geometry-aware peptide binding site prediction}

GeoPep takes peptide and protein sequences as input and outputs residue-wise binding probabilities, representing the confidence that each residue participates in peptide-protein interactions (Figure 1A). The architecture combines transfer learning from ESM3, a state-of-the-art multimodal protein foundation model trained on billions of sequences and structures to jointly reason over protein sequence, structure, and function, with sequential Kolmogorov-Arnold Network (KAN) modules. During training, GeoPep incorporates distance-based constraints to enforce spatial coherence in binding predictions (Figure 1B).

\begin{figure*}[t]
    \centering
    \includegraphics[width=\textwidth]{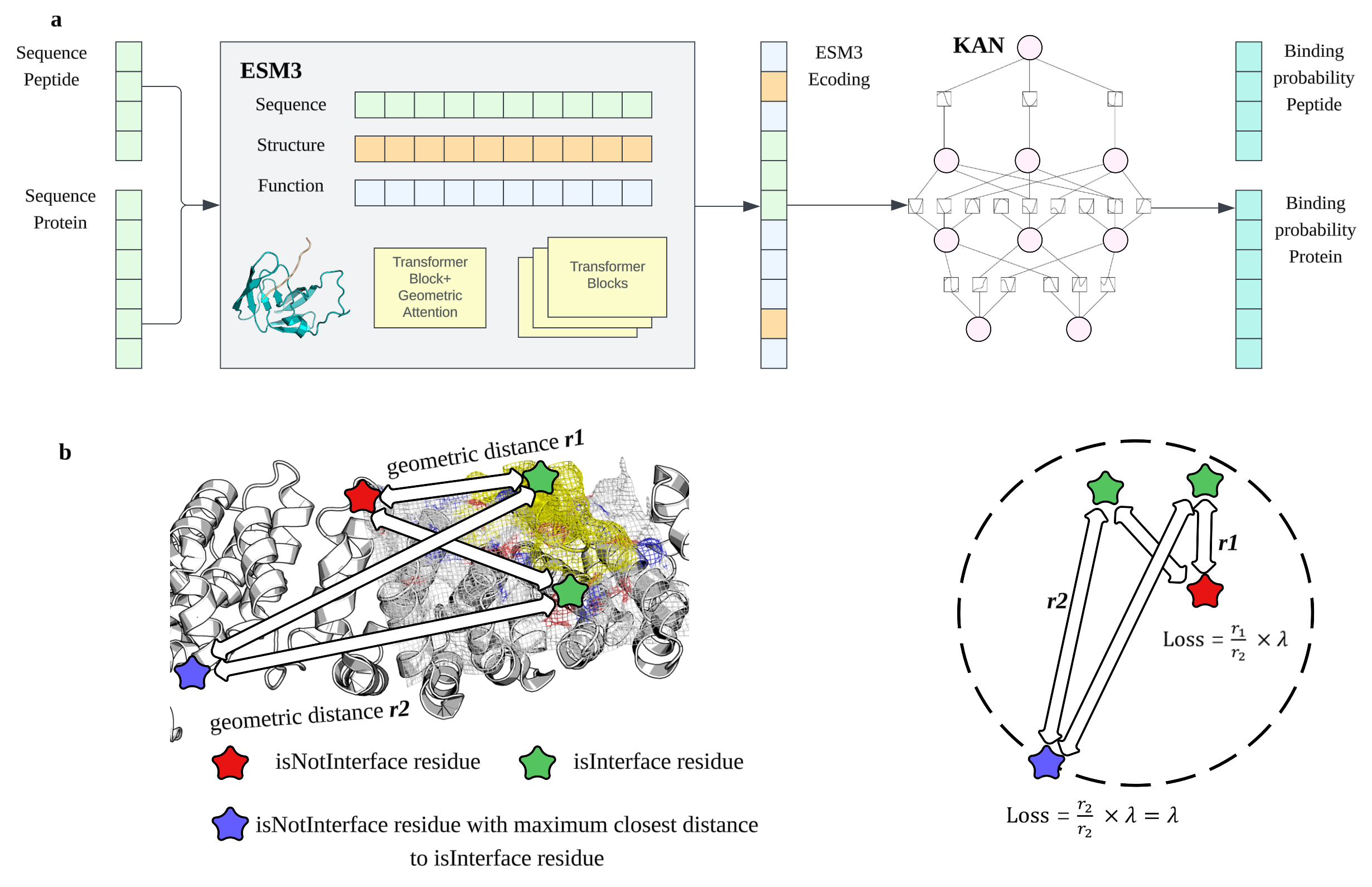}
    \caption{\textbf{a} GeoPep architecture combining ESM3 transfer learning with Kolmogorov-Arnold Network (KAN) modules for peptide binding site prediction. Peptide and protein sequences are processed through ESM3's multimodal encoder, which integrates sequence, structural, and functional information via transformer blocks with geometric attention mechanisms. The resulting ESM3 embeddings are passed to KAN layers that employ learnable B-spline activation functions to model complex nonlinear binding patterns, outputting residue-level binding probabilities. \textbf{b} Distance-based geometric loss function enforcing spatial consistency in binding site predictions. For each non-interface residue incorrectly predicted as a binding site (false positive), $r_1$ represents the minimum 3D distance to the nearest true interface residue, and $r_2$ represents the maximum such distance across all false positives in the complex. The geometric loss is computed as $(r_1/r_2) \times \lambda$, where $\lambda$ is the regularization weight. This distance-normalized penalty ensures that false positives farther from true binding sites incur larger penalties, guiding the model toward spatially coherent predictions clustered around verified interface residues. Detailed mathematical formulations are provided in Methods.}
    
    \label{fig:enter-label}
\end{figure*}

In GeoPep, ESM3 embeddings, which capture evolutionary and structural knowledge through geometric attention mechanisms that model three-dimensional spatial relationships among amino acids, are fed into sequential KAN modules. Unlike conventional multilayer perceptrons with fixed activation functions, KANs employ learnable activation functions on network edges, providing a parameter-efficient architecture that reduces computational overhead during fine-tuning of large foundation models. This efficiency is particularly advantageous for peptide binding prediction, where training on limited peptide–protein datasets benefits from reduced parameter requirements without sacrificing expressiveness. 

Distance-based loss functions are implemented to enforce spatial coherence of the predicted binding surface. These geometric constraints penalize configurations where residues assigned high binding confidence are geometrically distant from the peptide. This regularization improves localization fidelity and reduces reliance on large labeled datasets by injecting physically meaningful structures into the learning objective. The resulting predictions form contiguous, spatially plausible binding patches on the receptor surface, consistent with peptide–protein interaction topologies.

\subsection{Comparative Analysis of ESM2 and ESM3 Foundation Models in GeoPep}

ESM2 and ESM3 represent successive generations of protein foundation models with distinct capabilities. ESM2 is a sequence-only transformer trained via masked language modeling, lacking structural context. In contrast, ESM3 integrates sequence, structure, and functional information within a 98-billion parameter architecture, enabling modeling of geometric relationships essential for peptide-protein binding prediction.

To quantify the impact of structural signals on peptide binding prediction, we implemented GeoPep with both ESM2 and ESM3 as backbone models, maintaining identical downstream components, including KAN modules and distance-based loss functions. Both variants were fine-tuned on the same training datasets, augmented with 5000 negative samples, under identical hyperparameter configurations to ensure fair comparison. Model performance was evaluated on a validation dataset comprising diverse peptide-protein complexes with experimentally validated binding affinities and corresponding negative samples.

\begin{figure*}[!t]
    \centering
    \includegraphics[width=\textwidth]{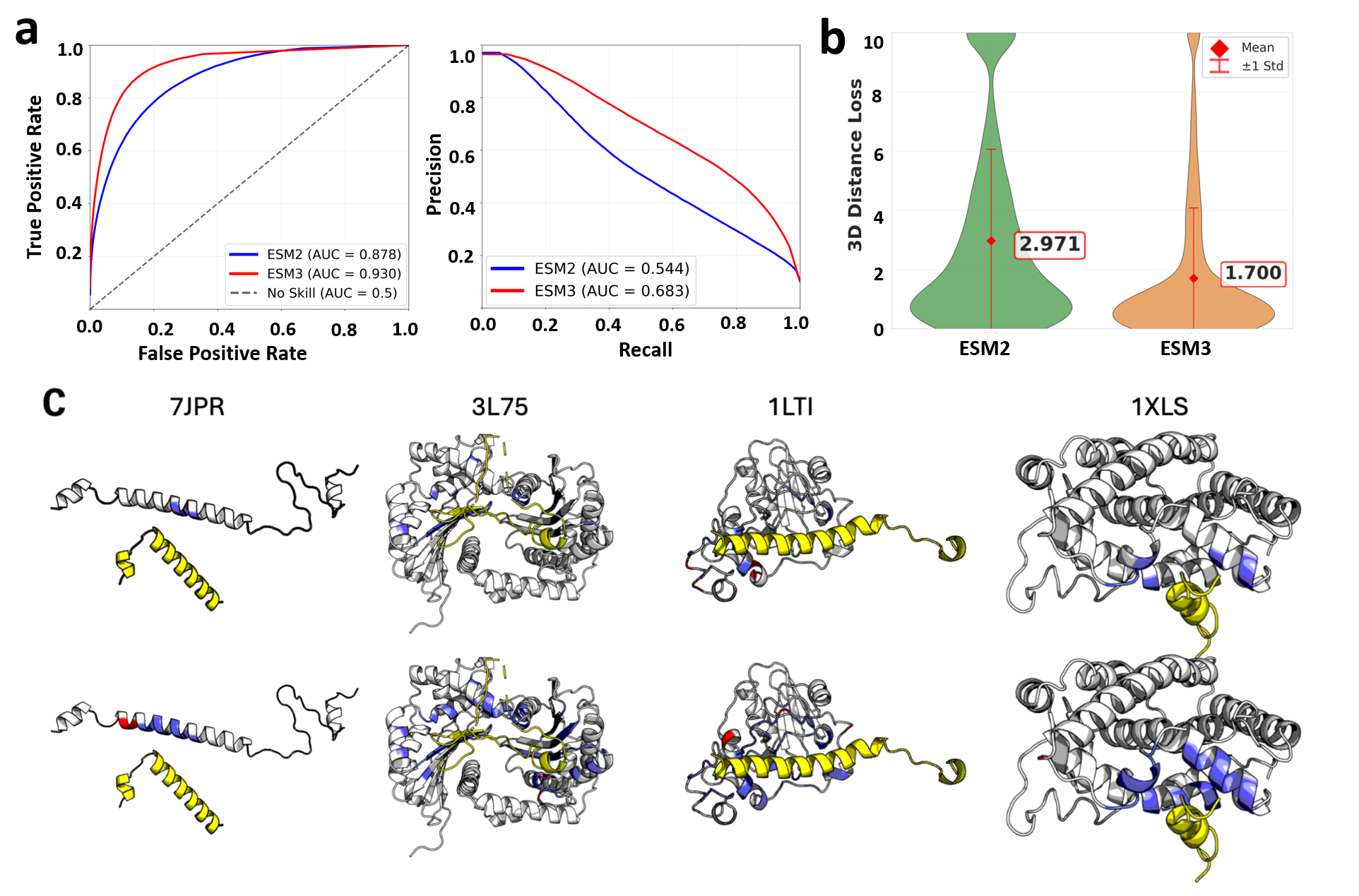}
    \caption{\textbf{a} ROC and precision-recall curves comparing GeoPep performance using ESM2 versus ESM3 backbones. \textbf{b} Distributions of 3D distance-based loss for ESM2 and ESM3, with calculation methodology detailed in Methods. \textbf{c} Binding site prediction visualizations for ESM2 (top) and ESM3 (bottom) on representative peptide-protein complexes (PDB IDs: \textbf{7JRP}, \textbf{3L75}, \textbf{1LTI}, and \textbf{1XLS}). Peptides are colored \textbf{yellow}, and true positives residues are highlighted in blue and false positives in red.}
    \label{fig:2}
\end{figure*}

The comparative evaluation demonstrates the superior performance of ESM3-GeoPep in all metrics (Figure~\ref{fig:2}). ESM3-GeoPep exhibits higher discriminative ability (ROC) and superior precision–recall characteristics relative to ESM2-GeoPep. Improvements are particularly pronounced in the high-precision regime relevant for identifying binding sites, indicating that structural priors encoded in ESM3 embeddings facilitate more selective identification of true binding residues under class imbalance. Analysis of the 3D distance-based metric demonstrates that ESM3-GeoPep attains lower mean distance errors between high-confidence residues and the peptide, consistent with the hypothesis that multimodal structural representations improve spatial localization of interaction hotspots. Moreover, the geometric loss landscape stabilizes earlier during training for ESM3-GeoPep, suggesting more faithful alignment between learned saliency and the physical contact manifold.

\subsection{Kolmogorov-Arnold Networks enhance training efficiency without sacrificing accuracy}


To evaluate the architectural contribution of KAN layers, we compared GeoPep variants using KAN modules versus traditional MLP modules while controlling for datasets (including negative samples) and training hyperparameters (Figure 3). KAN-based models converged faster and achieved superior performance across all evaluation metrics. Specifically, the KAN variant reached optimal performance in significantly fewer training epochs while delivering higher classification accuracy and improved geometric consistency. These results demonstrate that KAN's learnable activation functions provide advantages over fixed-activation MLPs in both training efficiency and final model quality for peptide binding site prediction.


\subsection{Distance-based geometric constraints improve prediction accuracy and training stability}

In addition to the standard cross-entropy loss for residue-level classification, GeoPep incorporates a distance-based geometric regularizer to enforce spatial coherence in predicted binding sites. While cross-entropy loss minimizes prediction errors for individual residues, it does not capture the spatial relationships among binding residues. The geometric loss penalizes configurations that violate known distance constraints, encouraging the model to learn spatially coherent binding patterns rather than isolated residue classification. This combined loss function balances classification accuracy with geometric plausibility, guiding the model toward predictions that are both statistically accurate and structurally realistic.

\begin{figure*}[!t]
    \centering
    \includegraphics[width=\textwidth]{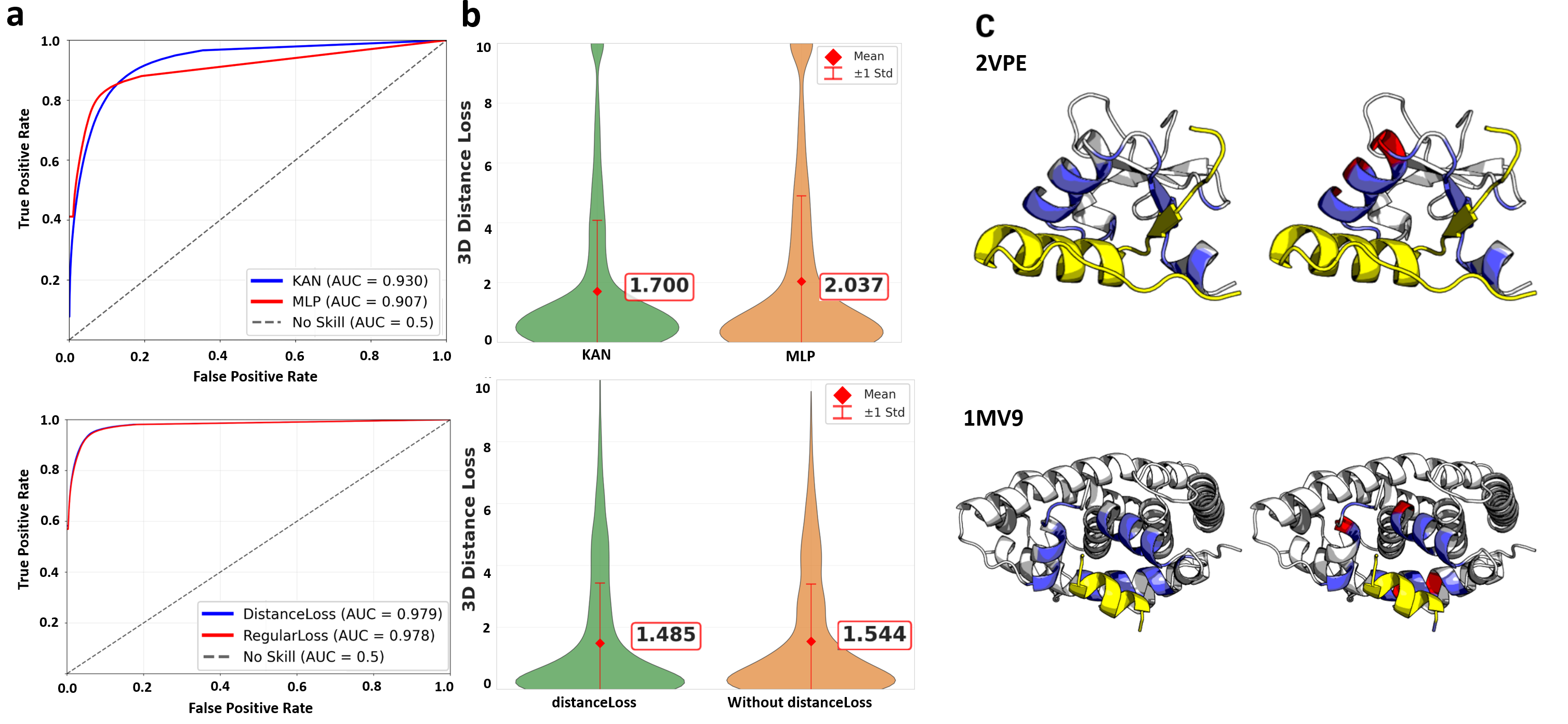}
    \caption{\textbf{a} ROC curves comparing KAN versus MLP architectures (top) and distance-based versus standard loss functions (bottom) in GeoPep. For the distance-based loss comparison, a sequence-level window size of 1 is used to expand ground-truth labels, providing a relaxed evaluation criterion. \textbf{b} Distributions of 3D distance-based loss for architectural and regularization comparisons, with calculation methodology detailed in Methods. \textbf{c} Representative binding site prediction visualizations for different model configurations. The top row shows predictions from KAN (left) and MLP (right) for PDB 2VPE, while the bottom row compares distance-based loss (left) and standard loss (right) configurations for PDB 1MV9.}
    \label{fig:3}
\end{figure*}

To evaluate its impact, we compared GeoPep variants trained with and without geometric regularization on identical datasets (Figure 3). Both approaches achieved similar ROC performance, indicating minimal impact on overall classification accuracy. However, models trained with geometric constraints exhibited lower mean distance errors and improved spatial continuity of predicted binding patches. These results indicate that while distance-based geometric constraints provide modest improvements in classification metrics, they substantially enhance spatial coherence and training stability, critical for applications requiring accurate geometric predictions.


\subsection{GeoPep Outperforms Existing Binding-Site Prediction Methods}

We benchmarked GeoPep against leading binding site prediction methods, including PepNN~\cite{abdin2022pepnn}, ScanNet~\cite{tubiana2022scannet}, and PeSTo~\cite{krapp2023pesto} using an independent validation dataset without negative samples (Figure 4). GeoPep, leveraging ESM3 embeddings and KAN integration, achieved superior performance across all metrics (Table~\ref{tab:performance_comparison}). Unless otherwise specified, residues with predicted probability $p_i \ge 0.8$ were classified as binding sites, while those with $p_i < 0.8$ were considered non-binding. This threshold was used consistently across all models for metric computation. ScanNet and PeSTo, both protein-protein binding site predictors, demonstrated limited effectiveness for peptide-protein prediction tasks, with weaker discriminative capability and lower precision-recall scores. PepNN, designed for peptide-protein complexes, performed better than protein-protein methods but remained below GeoPep across all evaluation criteria. 

\begin{table}[htbp]
\centering
\caption{Performance comparison of peptide binding site prediction methods at probability threshold = 0.8}
\label{tab:performance_comparison}
\begin{tabular}{|l|c|c|c|c|}
\hline
\textbf{Model} & \textbf{Precision} & \textbf{Recall} & \textbf{F1-score} & \textbf{Accuracy} \\
\hline
GeoPep         & 0.8496             & 0.9567          & 0.90              & 0.9604            \\
\hline
PepNN          & 0.5684             & 0.5713          & 0.5699            & 0.919             \\
\hline
PeSTo          & 0.3557             & 0.6436          & 0.4582            & 0.881             \\
\hline
ScanNet        & 0.3707             & 0.3146          & 0.3404            & 0.9205            \\
\hline
\end{tabular}
\end{table}

GeoPep outperformed all baseline methods in ROC and precision-recall metrics, and and demonstrated superior true positive volume ratio, indicating more continuous and spatially coherent binding surface predictions. Furthermore, 3D distance loss analysis confirmed GeoPep's geometric advantage, achieving the lowest mean distance error among all methods tested.  Collectively, these results validate that combining multimodal transfer learning from ESM3 with parameter-efficient KAN architectures and geometry-aware loss functions enables robust modeling of peptide–protein interactions, setting a new benchmark for binding site prediction.


\begin{figure*}[!t]
    \centering
    \includegraphics[width=\textwidth]{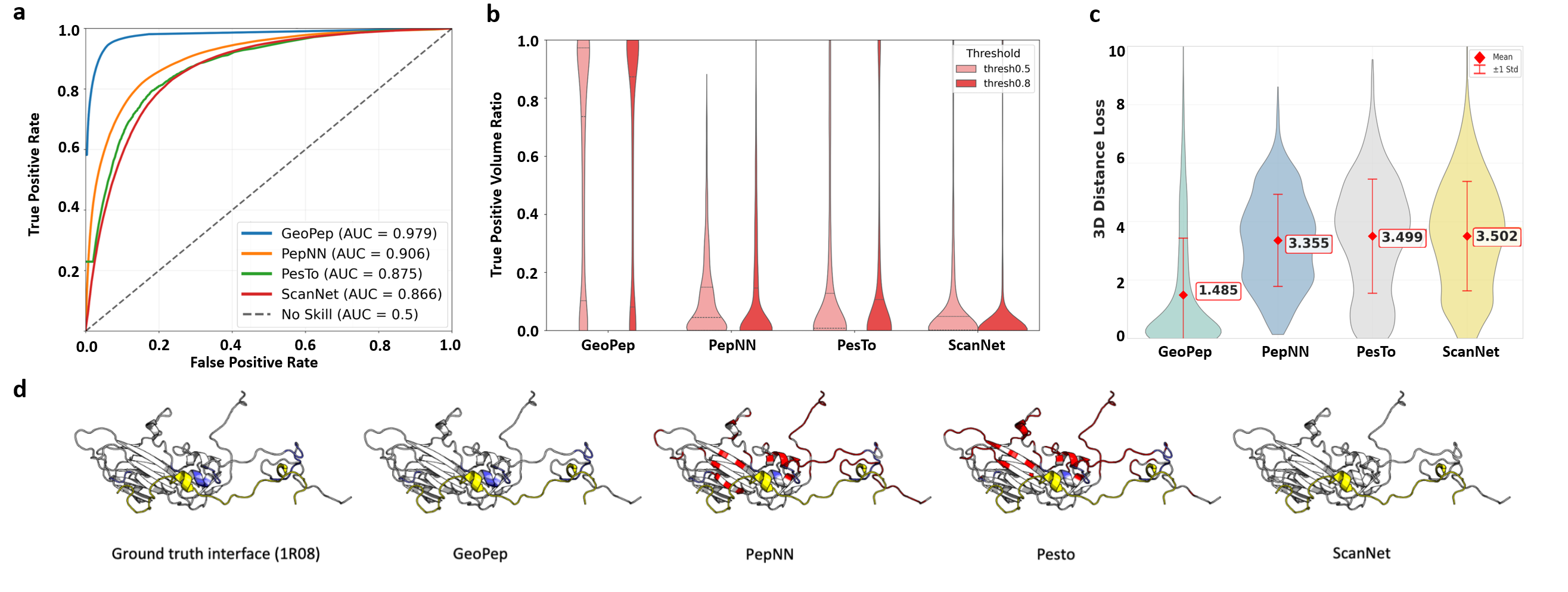}
    \caption{\textbf{a} ROC curves comparing GeoPep with baseline methods (PepNN, PesTo, ScanNet) computed using a sequence-level window size of 1 to expand ground-truth labels for a relaxed evaluation criterion. \textbf{b} True positive volume ratio (TPVR) analysis at two thresholds (0.5 and 0.8), calculated as the ratio of true positive volume to total predicted volume; higher ratios indicate more accurate and contiguous binding surface predictions. \textbf{c} Distributions of 3D distance-based loss across methods, with calculation methodology detailed in Methods. Lower values correspond to better geometric accuracy in binding site predictions. \textbf{d} Visualization of peptide–protein interface predictions for the 1R08 complex. Ground truth interface residues are shown in the leftmost panel. For predictions by GeoPep, PepNN, Pesto, and ScanNet, peptides are colored yellow, true positives are highlighted in blue, and false positives in red.}
    \label{fig:4}
\end{figure*}

\subsection{Structural evaluation}
To assess the performance of GeoPep in localizing peptide--protein interfaces without explicit structural input, we examined three representative complexes from the test set, ordered by increasing interface-residue count. As shown in Fig.~5a, \textit{Left (6U3O)} depicts a TCR--HLA-DQ2 complex presenting a microbial mimic peptide. The peptide adopts an extended register along the MHC-II peptide-binding groove (a $\beta$-sheet platform flanked by two $\alpha$-helices), and GeoPep correctly recovers the slender peptide--groove contact. \textit{Middle (3DNO)} shows HIV-1 gp120 bound to CD4; despite a solvent-exposed, loop-dominated interface---typically disfavored by physics-based scoring due to desolvation and configurational-entropy penalties---the model accurately pinpoints the gp120--CD4 contact region. \textit{Right (1A1M)} presents HLA-B*53:01 with a linear peptide lodged in the canonical MHC-I groove, for which the larger interface is correctly assigned.

\begin{figure*}[!t]
    \centering
    \includegraphics[width=\textwidth]{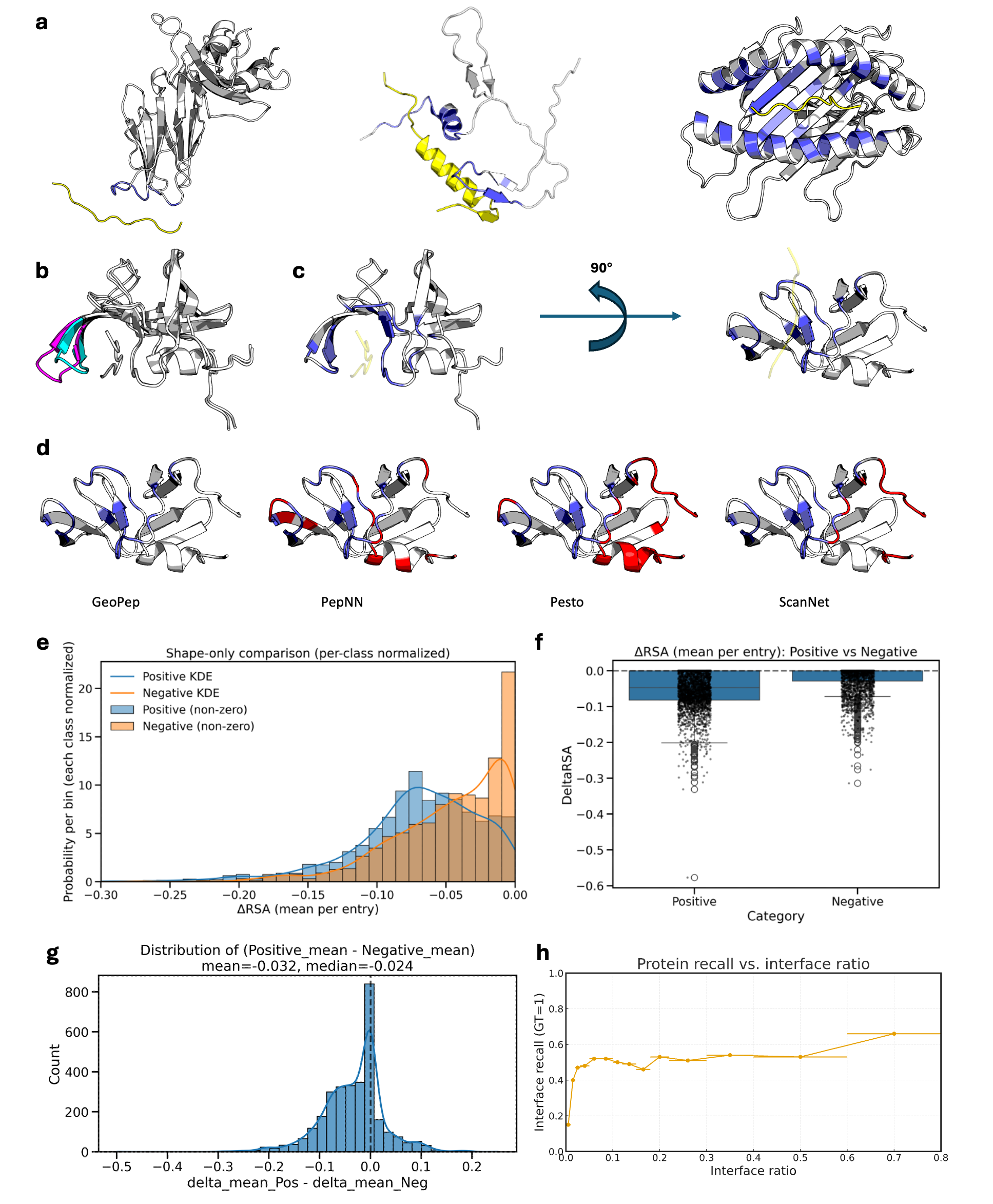}
    \caption{(a) Case studies of peptide--protein interface prediction with 100\% residue-level accuracy; peptides are shown in yellow and correctly predicted interface residues in slate. Left to right: 6U3O, 3DNO, 1A1M (interface-residue count increases from low to moderate to high). (b) Overlay of the apo (2PC0) and holo (2NXL) states of HIV protease illustrating conformational change upon peptide binding. (c) Holo state rotated by 90$^\circ$ about the indicated axis for structural clarity. (d) GeoPep predictions on the same target alongside three baselines (PepNN, Pesto, and ScanNet). (e) Per-class normalized histograms and kernel density estimates (KDEs) for Positive (blue) and Negative (orange) predictions, highlighting $\Delta$RSA distributions. (f) Distributions of per-entry $\Delta$RSA for Positive vs.\ Negative predictions. (g) Distribution of per-entry differences between the mean $\Delta$RSA of residues predicted as Positive and those predicted as Negative. (h) Protein-level interface recall plotted against interface ratio.}

    \label{fig:placeholder}
\end{figure*}

We next evaluated GeoPep’s performance in scenarios where peptide binding is coupled to substantial protein conformational changes. As shown in Fig. 5b, for the apo–holo states of HIV protease (PDBs: 2PC0 for apo; 2NXL for peptide-bound), a mobile loop (residues 45–55; apo in magenta, holo in cyan) undergoes a pronounced rearrangement upon peptide binding, transitioning from an open to a closed conformation that stabilizes the peptide interaction. Despite the absence of a pre-formed pocket in the apo state, the \textbf{structure-less GeoPep} model correctly identifies this loop as interfacial, demonstrating that the model is not restricted to rigid, preorganized sites but accommodates binding-induced structural rearrangements that often confound physics-based scoring functions.

On this induced-fit case, GeoPep was compared to established baselines (e.g., PepNN, Pesto, and ScanNet) (Fig. 5d). GeoPep concentrates predictions on the loop that closes upon binding, whereas baselines either miss portions of the contact band or disperse signal over neighboring solvent-exposed regions. 

To test whether predictions align with underlying biophysical principles, we examined residue-level solvent accessibility changes ($\Delta$RSA). A defining property of true interface residues is burial upon binding. As shown in Fig. 5e, residues predicted as Positives exhibit a clear shift toward more negative $\Delta$RSA values, while Negatives cluster near zero, indicating minimal change in solvent exposure. At the complex level (Fig. 5f), per-entry means of Positive and Negative residues display a consistent separation across 3,127 complexes, with Positives exhibiting markedly stronger burial (mean = $-0.0526$, median = $-0.0471$) than Negatives (mean = $-0.0202$, median $\approx$ 0.0). The paired differences (Positive $-$ Negative) for each complex form a unimodal distribution centered below zero (mean = $-0.0325$, median = $-0.0237$; Fig. 5g). A one-sided Wilcoxon signed-rank test (Pratt method) confirmed the significance of this shift (W = 922,715, p = $8.8 \times 10^{-184}$) with a large effect size (r = 0.62). Bootstrap 95\% confidence intervals excluded zero (mean difference [$-0.0346$, $-0.0304$]; median [$-0.0275$, $-0.0207$]). These results suggest that residues predicted as Positives consistently undergo stronger burial upon binding, indicating that GeoPep captures a core biophysical hallmark of peptide--protein binding.

As shown in Fig. 5h, recall on protein interface residues increases monotonically with interface size. Small or sparse interfaces yield low recall, reflecting weaker sequence–structural cues and higher background noise. In contrast, recall stabilizes around 0.5–0.6 for medium-to-large interfaces and occasionally exceeds 0.6 when the interface ratio surpasses 0.4. This trend suggests that the model preferentially captures larger, contiguous interaction patterns, consistent with enhanced learnable signals.

 For the induced-fit case (PDB: 2NXL), GeoPep's performance parallels the statistical trends in Fig. 5e–g and recall dependence in Fig. 5h: interfaces exhibiting larger binding-induced burial ($\lvert \Delta \mathrm{RSA} \rvert$), more contiguous contact patches, and higher interface ratios are recovered more reliably, whereas loop-dominated or low-burial surfaces tend to be under- or over-predicted by pocket-biased models. By not presupposing a pre-formed pocket, GeoPep remains effective on induced-fit interfaces that typically degrade the accuracy of structure- or geometry-based predictors.

\section{Discussion}

Our results demonstrate that GeoPep effectively leverages multimodal protein foundation models for peptide-protein binding site prediction through strategic architectural and training choices. The superior performance of ESM3 over ESM2 confirms that structural information encoded in multimodal foundation models provides critical advantages for capturing the geometric constraints inherent in peptide-protein interactions. The integration of distance-based geometric losses further validates this approach, as evidenced by improved spatial coherence in binding predictions compared to models without geometric constraints. This finding underscores the importance of incorporating three-dimensional structural information into sequence-based models, particularly for tasks requiring spatial understanding of molecular interactions. By enforcing geometric consistency, the regularization ensures that predicted binding sites maintain realistic three-dimensional relationships, addressing a key limitation of purely sequence-based approaches.

The adoption of Kolmogorov-Arnold Networks (KANs) as downstream predictors demonstrates a parameter-efficient strategy for fine-tuning large foundation models. KAN-based architectures achieve comparable or superior performance to traditional MLPs while reducing computational overhead and accelerating convergence. This suggests promising directions for efficient adaptation of large protein language models to specialized tasks. Overall, GeoPep's combination of multimodal representations, parameter-efficient architecture and geometric regularization outperforms both specialized peptide-protein methods and general protein-protein interface predictors. These results highlight the effectiveness of integrating structural knowledge with transfer learning for specialized molecular interaction tasks.

Several promising avenues emerge for extending this work beyond binary binding site classification. The distance-based geometric constraints introduced here could be adapted for predicting two-dimensional contact maps or generating three-dimensional peptide-protein complex structures, as the spatial relationships encoded in the distance-based loss function provide a natural framework for structure generation. However, the current work faces notable limitations that should be addressed in future studies. The peptide-protein structural dataset remains relatively small compared to protein-protein interaction databases, limiting generalization across diverse binding modalities. 
Additionally, the current dataset lacks systematic classification and balanced sampling strategies, potentially introducing training biases toward specific interaction types or structural motifs. Future work should prioritize expanding datasets, developing standardized classification schemes and addressing data distribution imbalances during training. Finally, extending GeoPep to binding affinity prediction rather than binary classification would significantly enhance its utility for therapeutic applications. Integration with drug discovery pipelines could further facilitate rational design of peptide inhibitors and modulators, bridging the gap between property prediction and generative modeling.

\section{Method}
\subsection{Datasets}

We constructed the dataset from Propedia v2.3, which aggregates peptide--protein complexes from the Protein Data Bank (PDB). Complexes were filtered to include peptides of length 2--50 residues and proteins resolved by X-ray crystallography at $\leq$2.5 \AA\ or NMR spectroscopy. Binding residues were defined as those within 6 \AA\ of any peptide atom. Additional filtering retained complexes with peptides exceeding 10 residues and protein chains less than 500 residues, yielding 32,290 peptide-protein chain combinations from 12,540 unique complexes. Each chain combination represents an individual peptide-protein interaction pair used for training. The dataset was randomly split into training (90\%, 29,061 instances) and validation (10\%, 3,229 instances) subsets.

\subsection{Input Preparation}
Input structures were processed using BioPython to extract atomic coordinates and sequence information from PDB files. Peptide sequences were padded to 50 residues and protein sequences to 500 residues to ensure consistent batch dimensions. Peptide-protein sequence pairs were tokenized using ESM3's sequence tokenizer.

\subsection{Model Architecture}

Input peptide-protein sequence pairs are processed through ESM3, a multimodal protein foundation model that encodes molecular representations via multi-head self-attention mechanisms across sequence, structural, and functional modalities. The resulting CLS token embedding, which represents the global complex state, is passed to a Kolmogorov-Arnold Network (KAN) stack. Each KAN layer implements learnable activation functions parameterized as B-spline basis functions with trainable coefficients, enabling enhanced modeling of complex binding relationships. The final layer outputs per-residue binding confidence scores ranging from 0 to 1.

\subsection{Training Objective}

The composite loss combines binary cross-entropy for residue-level classification with a distance-based geometric regularization term:
$$L_{\text{total}} = L_{\text{CE}} + \lambda L_{\text{struct}}$$

The cross-entropy loss for binary classification is:
$$L_{\text{CE}} = -\frac{1}{N}\sum_{i=1}^{N} \left[ y_i \log(p_i) + (1-y_i) \log(1-p_i) \right]$$
where $y_i \in \{0,1\}$ indicates whether residue $i$ is a true binding residue, $p_i$ is the predicted probability of residue $i$ being a binding residue, and $N$ is the total number of residues.

The structural regularization term penalizes false positives proportionally to both confidence and distance from true binding sites:
$$L_{\text{struct}} = \frac{1}{N}\sum_{i=1}^{N} L_{\text{struct}}(i)$$
$$L_{\text{struct}}(i) = \begin{cases} 
0 & \text{if } y_i = 1 \\
p_i \times \frac{d_{\text{3D}}(i)}{\max(d_{\text{3D}})} & \text{if } y_i = 0
\end{cases}$$
where $d_{\text{3D}}(i) = \min_{j \in I} \|\mathbf{r}_i - \mathbf{r}_j\|_2$ is the minimum 3D distance from residue $i$ to any true binding residue, $I$ denotes the set of true binding residues, $\max(d_{\text{3D}}) = \max_{i=1}^{N} d_{\text{3D}}(i)$ is the maximum 3D distance across all residues. $\lambda$ is a hyperparameter controlling the regularization strength. This regularization term enforces spatial coherence by discouraging high-confidence predictions far from true interfaces.

\subsection{True Positive Volume Ratio}

To evaluate binding surface continuity, we introduce the true positive volume ratio metric that quantifies the spatial coherence of predicted binding sites. The convex hull algorithm constructs the smallest convex polytope enclosing all input points, capturing the spatial extent of residue clusters. Let $R_{\text{TP}}$ denote true positive residues and $R_{\text{pred}}$ denote all predicted binding residues. The convex hull volumes are:

$$V_{\text{TP}} = \text{ConvexHull}(\{\mathbf{r}_i : i \in R_{\text{TP}}\})$$

$$V_{\text{pred}} = \text{ConvexHull}(\{\mathbf{r}_i : i \in R_{\text{pred}}\})$$

where $\mathbf{r}_i$ represents mass center of residue $i$, $V_{\text{TP}}$ is the volume enclosed by true positive residues, and $V_{\text{pred}}$ is the volume enclosed by all predicted residues. The true positive volume ratio is:

$$\text{TP Volume Ratio} = \frac{V_{\text{TP}}}{V_{\text{pred}}}$$

Since $R_{\text{TP}} \subseteq R_{\text{pred}}$, this ratio ranges from 0 to 1. Higher ratios indicate compact, continuous binding surfaces where false positives cluster near true positives. Lower ratios suggest spatially dispersed predictions with fragmented binding sites.

\subsection{Training}
The model is trained to predict peptide binding sites on protein surfaces. The GeoPep architecture was trained for 7 days on 8 NVIDIA RTX A5000 GPUs using PyTorch Lightning with FSDP (Fully Sharded Data Parallel) distributed training strategy. The training objective employs a composite loss function combining binary cross-entropy loss for residue-level binding classification and the distance-based regularization loss. Gradient descent with automatic mixed precision is used to optimize the 1.4 billion model parameters.

\subsection{Evaluation}

Our method was compared with PepNN, ScanNet, and PeSTo to assess performance on peptide-protein binding site prediction. PepNN is a specialized deep learning method designed for peptide-protein interface prediction, serving as the most relevant baseline for our task. ScanNet and PeSTo are state-of-the-art geometry-based deep learning methods developed for protein-protein interface prediction, included to evaluate the transferability of protein-protein models to peptide binding tasks. The benchmarking of GeoPep was performed using structures from our validation dataset exclusively, comprising 3,229 peptide-protein instances derived from unique structural complexes in the Propedia v2.3 database.

\end{document}